\documentclass[
 reprint,
superscriptaddress,
 amsmath,amssymb,
 aps,
]{revtex4-2}

\usepackage{graphicx}% Include figure files
\usepackage{dcolumn}% Align table columns on decimal point
\usepackage{bm}% bold math
\usepackage{braket}
\usepackage{orcidlink}

\usepackage{enumitem} % Para listas organizadas
\newcommand{\be}{\begin{equation}}
\newcommand{\ee}{\end{equation}}
\newcommand{\bea}{\begin{eqnarray}}
\newcommand{\eea}{\end{eqnarray}}
\newcommand{\ave}[1]{\left\langle #1   \right\rangle }

\newcommand{\x}{^{\dagger}}

\usepackage{hyperref} % Para enlaces dentro del documento
\hypersetup{
    colorlinks,
    linkcolor={blue!80!blue},
    citecolor={blue!80!blue},
    urlcolor={blue!80!blue}
}
\newcommand{\mP}{\mathcal P}
\newcommand{\mQ}{\mathcal Q}
\newcommand{\mT}{\mathcal T}
\newcommand{\mE}{\mathcal E}
\newcommand{\mU}{\mathcal U}
\newcommand{\mL}{\mathcal L}
\newcommand{\mK}{\mathcal K}

\usepackage{physics}
\usepackage{float} % Agrega este paquete en el preámbulo

\usepackage{listings}
\begin{document}

\preprint{APS/123-QED}

\title{The Transfer Tensor Method: an Analytical Study Case}% Force line breaks with \\

\author{Marcel Morillas-Rozas\,\orcidlink{0009-0005-4570-1016}}%
 \email{marcel.morillas@upct.es}
\affiliation{Research Group of Quantum Technologies, Universidad Politécnica de Cartagena member of European University of Technology EUT+, Cartagena E-30202, Spain}

\author{Alberto López-García\,\orcidlink{0009-0001-0023-9850}}%
\affiliation{Research Group of Quantum Technologies, Universidad Politécnica de Cartagena member of European University of Technology EUT+, Cartagena E-30202, Spain}

\author{Gonzalo Reina Rivero\,\orcidlink{0000-0003-4219-2306}}%
\affiliation{Research Group of Quantum Technologies, Universidad Politécnica de Cartagena member of European University of Technology EUT+, Cartagena E-30202, Spain}

\author{Jianshu Cao}%
\affiliation{Department of Chemistry, Massachusetts Institute of Technology, Cambridge, MA 02139, United States of America}

\author{Javier Cerrillo\ \orcidlink{0000-0001-8372-9953}}%
 \email{javier.cerrillo@upct.es}
\affiliation{Research Group of Quantum Technologies, Universidad Politécnica de Cartagena member of European University of Technology EUT+, Cartagena E-30202, Spain}

%\date{\today}

\begin{abstract}
The transfer tensor method (TTM) is a versatile tool for analyzing and propagating general open quantum systems. It captures in a compact manner all memory effects in a non-Markovian system through a straightforward transformation of a set of dynamical maps. Transfer tensors provide the exact convolutional propagator associated with a given time discretization over the past evolution of an open quantum system. Here we show that, for any finite time discretization, the memory kernel of the Nakajima Zwanzig equation deviates from the exact transfer tensors, although both converge in the continuous-time limit, as expected. We examine this behaviour in the context of an analytically solvable model: a two level atom resonant with a lossy cavity in the Jaynes Cummings limit. The atomic dynamics separate into two decoupled subspaces -- the coherence and the population difference. We derive exact expressions for the dynamical map, the transfer tensors and the memory kernel governing the coherence, and we relate them to their counterparts for the population difference. As a function of the ratio between the cavity loss rate and the atom-cavity coupling strength, we identify regions of enhanced non-Markovianity in which the dynamics can be described as fully Markovian for certain TTM time-step choices.
\end{abstract}

\maketitle

%\tableofcontents

\section{\label{sec:introduction}Introduction}
Transfer tensors (TT) are superoperators that facilitate the propagation and analysis of non-Markovian open quantum systems. They were originally introduced as a way to extract information from short--time trajectories and extrapolate the behaviour of the system to longer times \cite{Cerrillo2014}. This is particularly useful in situations where the computational cost increases rapidly with the target simulation time, as is often the case for open quantum systems strongly coupled to an environment and exhibiting non-Markovian behaviour \cite{deVega2017, BriandGeva2021}. Transfer tensors can be understood as a transformation of a set of dynamical maps describing the propagation of an open quantum system over a translational invariant time series \cite{Cerrillo2014,PollockModi2018}.

Since their introduction, TT have found a variety of applications as both a propagation and an analysis tool.  Depending on the technique used to calculate the dynamical map, they have been employed, for instance, to extend numerically exact simulations to longer times \cite{Rosenbach2016, WuCerrilloCao2024}, to account for initially correlated system--environment states \cite{Buser2017}, to analyze the applicability of discrete-time propagation schemes in open-system dynamics \cite{Cygorek2017}, and to characterize non-Markovian noise from reconstructed quantum processes \cite{Chen2020}. Their range of applicability has also been studied in detail \cite{Gelzinis2017}. Depending on the technique used to calculate the dynamical map, different brands of TT have emerged \cite{Makri2020, Wang2024}. In general terms, the transfer tensor method sheds light on the operational meaning of quantum Markovianity \cite{Pollock2018Markov} and multi-temporal memory effects can be described with a generalization of the transfer tensor formalism known as the process tensor \cite{Pollock2018ProcessTensor}.

These developments are naturally linked to the theory of generalized master equations. In particular, a great deal of effort has been put into the Nakajima-Zwanzig equation (NZE) \cite{Nakajima1958, Zwanzig1960} and practical strategies for building or discretizing its memory kernel \cite{ShiGeva2003, Kelly2016, BriandGeva2021, Makri2025, Peng2025}. Although TT are closely related to this line of research, they are conceptually different objects. They are defined through a discrete family of dynamical maps and, therefore, provide an exact discrete-time representation of the reduced dynamics on a given time series \cite{PollockModi2018}. \textcolor{black}{In the continuous time limit, TTs can be connected to the NZE memory kernel. However, for finite timesteps the TTs are a completely different mathematical object than the memory kernel of the NZE. TTM is, by definition, an exact method to extract a propagator in a non-Markovian setting; and the only error introduced by this method comes from the truncation on the number of dynamical maps used \cite{Peng2025}. In contrast, in addition to the error corresponding to the truncation of the memory time, the discretization of the NZE memory kernel also introduces an error. \cite{Makri2025, Peng2025}.}

\textcolor{black}{This article aims to address the widespread notion that the TTM is a discretization of the NZE \cite{Makri2025, Peng2025}. In contrast, the input required to extract the TTs is the exact dynamical maps, not discretized versions of a memory kernel that are not unique and, in general, more difficult to compute than dynamical maps. The formalism used to generate the dynamical maps plays no role in the definition of the TTs. The dynamical maps can be obtained from stochastic methods, hierarchical equations of motion, path-integral techniques, or any other numerically exact approach.}

In this paper, we analyze the mathematical construction of TTs and distinguish it from discretizations of the memory kernel. We resort to a theoretical model that allows us to derive transfer tensors fully analytically. Their study provides insights into the dynamics such as the existence of Markovian time-discretizations in otherwise fully non-Markovian regimes. This analytical model will assist further future analysis on the properties of TTM. This manuscript is structured as follows: In Section~\ref{sec:intro_TTM} we introduce and compare the transfer tensors and the memory kernel of the Nakajima-Zwanzig equation, and we present the atom-in-a-cavity model used in Section~\ref{sec:JCM}. Section~\ref{sec:coherences} and \ref{sec:populations} are devoted to the analytical derivation and study of the memory kernels, dynamical maps, and transfer tensors of the system. Finally, we draw conclusions in Section~\ref{sec:conclusions}.

\section{The Transfer Tensors and the Memory Kernel}\label{sec:intro_TTM}
Given a homogeneous discretization of time $t_k=k\delta t$, a family of dynamical maps $\lbrace\mE_k\rbrace$ may be defined such that the density matrix of the open quantum system evolves as $\rho(t_k)=\mE_k\rho(0)$. The transfer tensors are then constructed iteratively from the dynamical maps via   
\be
\mT_k=\mE_k-\sum_{j=1}^k\mT_j \mE_{k-j},
\ee
which rewrites the propagation in the exact convolutional form
\be
\rho(t_k)=\sum_{j=1}^k\mT_j\rho(t_{k-j}).
\ee
An especially insightful object in this hierarchy is the second transfer tensor $\mathcal{T}_2$. It quantifies the deviation incurred when attempting to decompose the two-step dynamical map $\mathcal{E}_2$ into the Markovian composition of two one-step maps, i.e.\ the error in approximating $\mathcal{E}_2 \approx \mathcal{E}_1^2$. In this sense, $\mathcal{T}_2$ provides a direct measure of the memory effects generated between the first and second time steps. This connects naturally to established notions of quantum non-Markovianity, where memory effects are associated either with the breakdown of CP-divisibility~\cite{Rivas2010} or with the revival of information flow as quantified, for instance, by trace-distance backflow~\cite{Breuer2009}. The magnitude of $\mathcal{T}_2$ may therefore be interpreted as a discrete-time analogue of these criteria, capturing the first departure from Markovian composition at finite resolution. 

\textcolor{black}{It is instructive to interpret the transfer tensor method from the perspective of a Markovian embedding. Although the reduced dynamics of the open quantum system is generally non-Markovian, it can always be represented as the projection of a Markovian evolution acting on a sufficiently enlarged Hilbert space. If we define the time-step propagator of the enlarged space as $\mathcal{U}(\delta t)$ and $\mathcal{P}$ as the projector onto the physical system of interest, the reduced dynamical maps in a time homogeneous evolution, i.e. $\mU(k\delta t) = \mU^k$ read}
\begin{equation}\label{eq:dynamical_maps_general}
    \textcolor{black}{\mathcal{E}_k = \mP\mU^k\mP.} 
\end{equation}
Within this representation, the $k$-th transfer tensor takes the compact form \textcolor{black}{(see Appendix.~\ref{sec:Tk_appendix})}
\begin{equation}\label{eq:Tk}
\mT_k=\mP\mU\mQ\left(\mQ\mU\mQ\right)^{k-2}\mQ\mU\mP,
\end{equation}
where $\mQ=1-\mP$ denotes the complementary projector.
The memory kernel (MK) of the Nakajima-Zwanzig formalism is defined differently. Starting from the first order differential equation satisfied by $\mU$
\be
\frac{d}{dt}\mU=\mL\mU,
\ee
the action of $\mP$ yields a closed integrodifferential equation for the dynamical map
\be
\frac{d}{dt}\mE(t)=\mP\mL\mP\mE(t)+\int_0^t\mK (t-s)\mE(s)ds,
\ee
with memory kernel
\be
\mK (t)=\mP\mL\mQ\exp\left(\mQ\mL\mQ t\right)\mQ\mL\mP.
\ee
Although the transfer tensor and the memory kernel differ significantly at any finite time discretization, both converge in the continuous-time limit
\be
\lim_{\delta t \to 0}\frac{\mT_k\left(\delta t\right)}{(\delta t)^2}=\mK(t),
\label{eq:lim}
\ee
as shown in Appendix~\ref{sec:TTM_K_lim}. Hence, they cannot coincide at finite $\delta t$ values. \textcolor{black}{Although the discretization of the memory kernel is not unique, with different schemes leading to different finite-step kernels, the transfer tensors are obtained through an exact transformation of the exact dynamical maps. Therefore, TTM constitutes the exact discrete-time description of the reduced dynamics, whereas any finite-step discretization of the continuous-time memory kernel is necessarily affected by discretization error.}

\section{The Jaynes-Cummings Model}\label{sec:JCM}

\begin{figure}
    \centering
    \includegraphics[trim = {3.5cm 2cm 3cm 3cm}, clip = True, width=\linewidth]{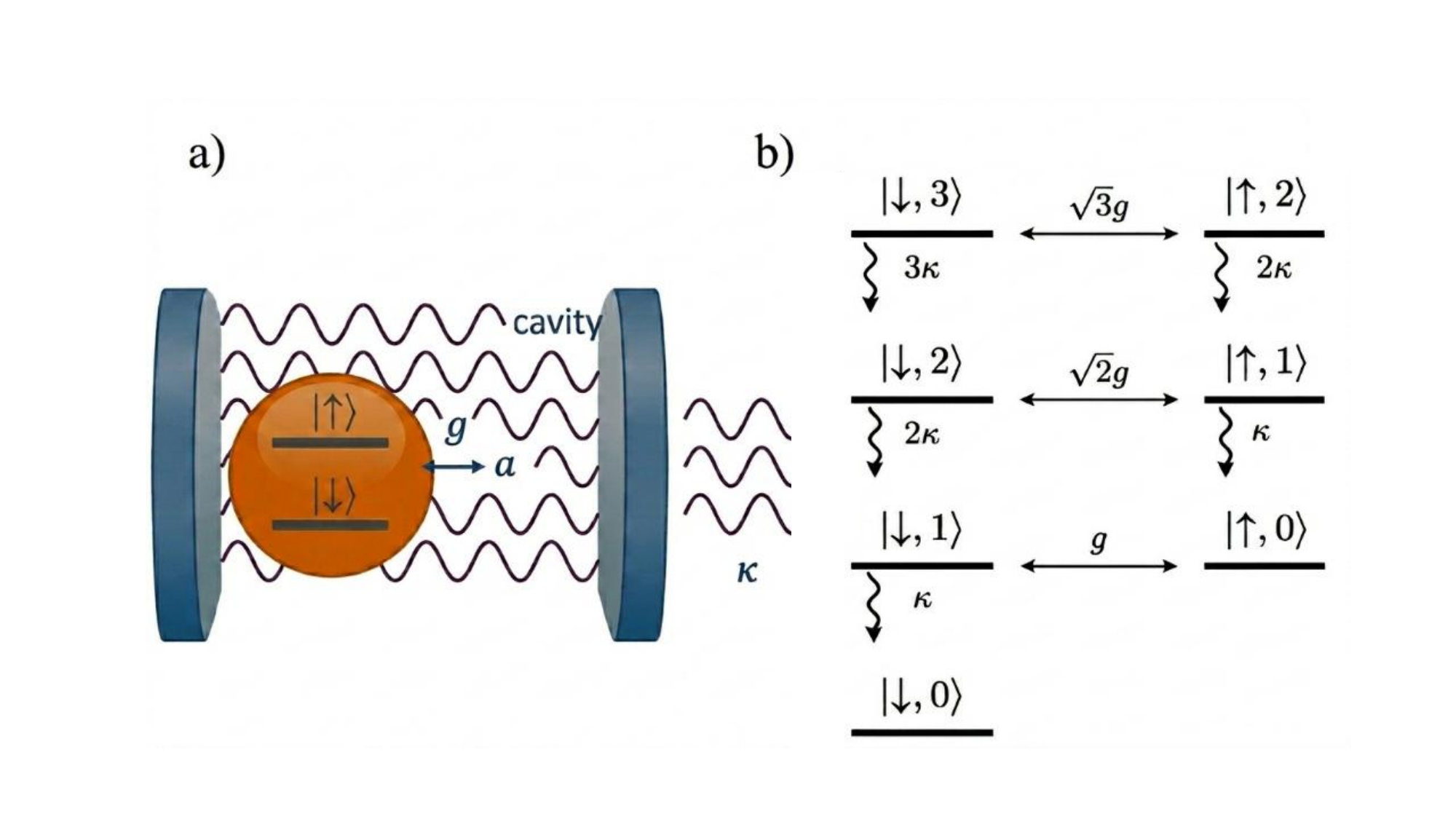}
    \caption{a) Diagram of the considered atom-cavity model. The atom is represented by two levels \textcolor{black}{$\ket \uparrow$} and \textcolor{black}{$\ket \downarrow$}. The cavity loses photons at a rate $\kappa$. Cavity and atom are coupled by a Jaynes-Cummings term of strength $g$. b) Level diagram representing the lowest states of the full Hilbert space, the strength of the couplings among themselves and the decay rates.}
    \label{fig:layout}
\end{figure}

In order to investigate the differences, let us consider an analytically solvable model consisting of a two-level atom of states \textcolor{black}{$\ket \uparrow$} and \textcolor{black}{$\ket \downarrow$} coupled to a single mode of a lossy cavity \textcolor{black}{with bosonic operators $\hat{a}$, $\hat{a}^\dagger$} in the Jaynes Cummings limit (see Fig.\ref{fig:layout}a). \textcolor{black}{The Hamiltonian of the system reads}
\begin{equation}
    \textcolor{black}{\hat{H} = \omega\hat{\sigma}_z + \omega_c\hat{a}^\dagger\hat{a} + g\hat{\sigma}_x(\hat{a}^\dagger + \hat{a}),}
\end{equation}
\textcolor{black}{where $\omega$ is the two-level atom frequency, $\omega_c$ is the cavity's frequency, $g$ is the coupling between the atom and the cavity and $\hat{\sigma}_{x,z}$ are Pauli matrices. In the rotating frame with respect to $\hat{H}_0 = \omega\hat{\sigma}_z + \omega_c\hat{a}^\dagger\hat{a}$ and applying the rotating wave approximation with the resonance condition ($2\omega = \omega_c$), the Hamiltonian of the system reads}
\be
\hat{H}'=g\hat{\sigma}^+\hat{a}+\text{H.c},
\ee
where $\hat{\sigma}^+ = \ketbra{\uparrow}{\downarrow}$. The master equation that describes the Markovian temporal evolution of the system reads
\begin{equation}\label{eq:master_eq}
\frac{d}{dt}\hat{\mu}=\left[\hat{H}',\hat{\mu}\right]+\kappa \left(\hat{a}\hat{\mu} \hat{a}\x-\frac 1 2 \hat{a}\x \hat{a}\hat{\mu}-\frac 1 2 \hat{\mu} \hat{a}\x \hat{a}\right).
\end{equation}
For a cavity initialized in its stationary state $\ket{0}$, just three states contribute to the dynamics: $\ket{e}=\ket{\uparrow, 0}$, $\ket{i}=\ket{\downarrow, 1}$ and $\ket{g}=\ket{\downarrow, 0}$ (see Fig.~\ref{fig:layout}b). We may then define the populations $p_k=\ave{k|\hat{\mu}|k}$ and the coherences $c_{jk}=\ave{j|\hat{\mu}|k}=c^*_{kj}$. The dynamics of the system decouples into two subspaces: on the one hand, subspace $A$ involving $\left\{p_e,p_i,p_g,c_{ei}\right\}$ and, on the other hand, subspace $B$ containing $\left\{c_{eg},c_{ig}\right\}$. Based on this decomposition, we may express the Liouvillian governing the dynamics as $\mL=\mL_A \oplus \mL_B$.

We now focus on the two level atom as our open quantum system of interest. This offers a minimal testing ground to analyze the effect that the strength of the coupling to the environment (represented by $g$) and the timescale of the bath (proportional to $\kappa^{-1}$) has on the Markovianity of the atom. The reduced density matrix of the system $\hat{\rho}$ contains elements $p_\uparrow=p_e$, $p_\downarrow=p_g+p_i$ and $c=c_{eg}$. The $A$-$B$ decomposition proves especially useful, as it enables us to derive the dynamical maps, transfer tensors, and memory kernels for both the population difference $\delta p=p_\uparrow-p_\downarrow$ and the coherence $c$ independently, and to express them as scalars rather than tensors.

\section{Coherence subspace}\label{sec:coherences}

For the smaller subspace $B$ the dynamical equations derived from Eq.~\eqref{eq:master_eq} for the coherences $\left\{c_{eg},c_{ig}\right\}$ read
\begin{equation}
    \left\lbrace\begin{array}{rcl}
        \dot c_{eg}&=&-ig c_{ig},\\
        \dot c_{ig}&=&-ig c_{eg}-\frac{\kappa}{2}c_{ig},
    \end{array}\right.
\end{equation}
which may be expressed in matrix from
$\dot{\mathbf{c}} = \mL_B \mathbf{c}$
with 
\be
\mathbf{c}=\begin{pmatrix}
c_{eg} \\
c_{ig}
\end{pmatrix};\qquad
 \mL_B =\begin{pmatrix}
0 & -i g \\
-i g & -{\kappa}/{2}
\end{pmatrix}.
\ee
The projection superoperators in this subspace become the matrices  $\mP_B = 1\oplus 0$ and $\mQ_B = 0\oplus 1$ which yield $\mQ_B\mL_B\mQ_B=-\kappa \mQ_B/2$ and $\mP_B\mL_B\mQ_B\mL_B\mP_B=-g^2\mP_B$, both elements required to calculate the memory kernel component for the coherence $c$
\be
\mK_c(t)=-g^2e^{-\kappa t/2}.
\ee

\begin{figure}
    \centering
    \includegraphics[width=\linewidth]{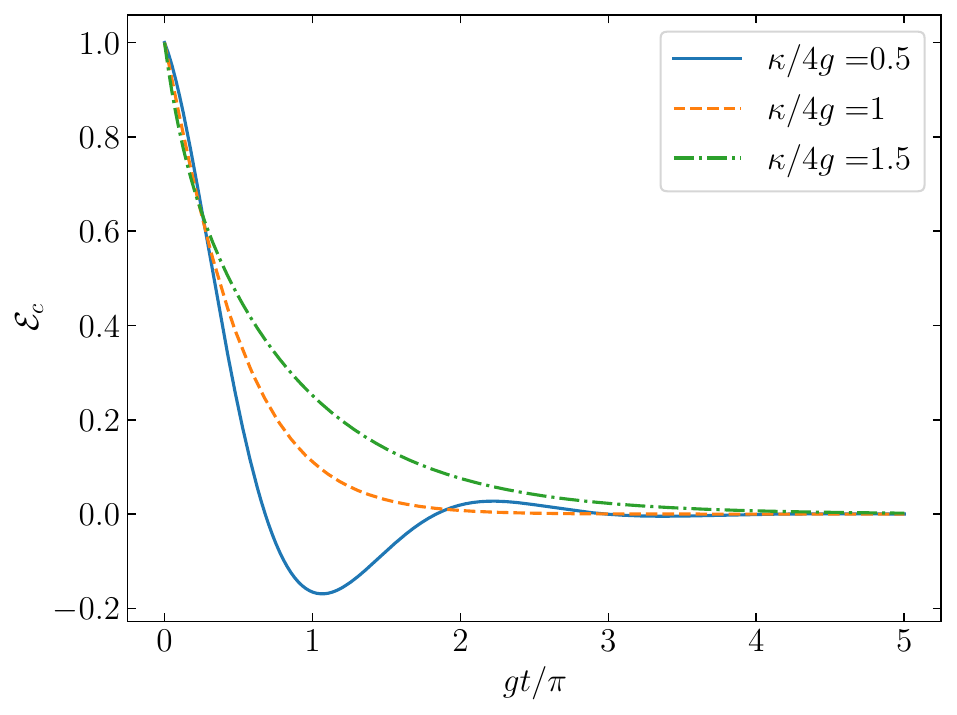}
    \caption{Temporal evolution of the coherence dynamical map $\mathcal{E}_c$ for different values of $\kappa/4g$, illustrating the underdamped (solid line), overdamped (dot-dashed line) and critically damped (dashed line) regimes.}
    \label{fig:3_regimes}
\end{figure}

\begin{figure}
    \centering
    \includegraphics[width = \linewidth]{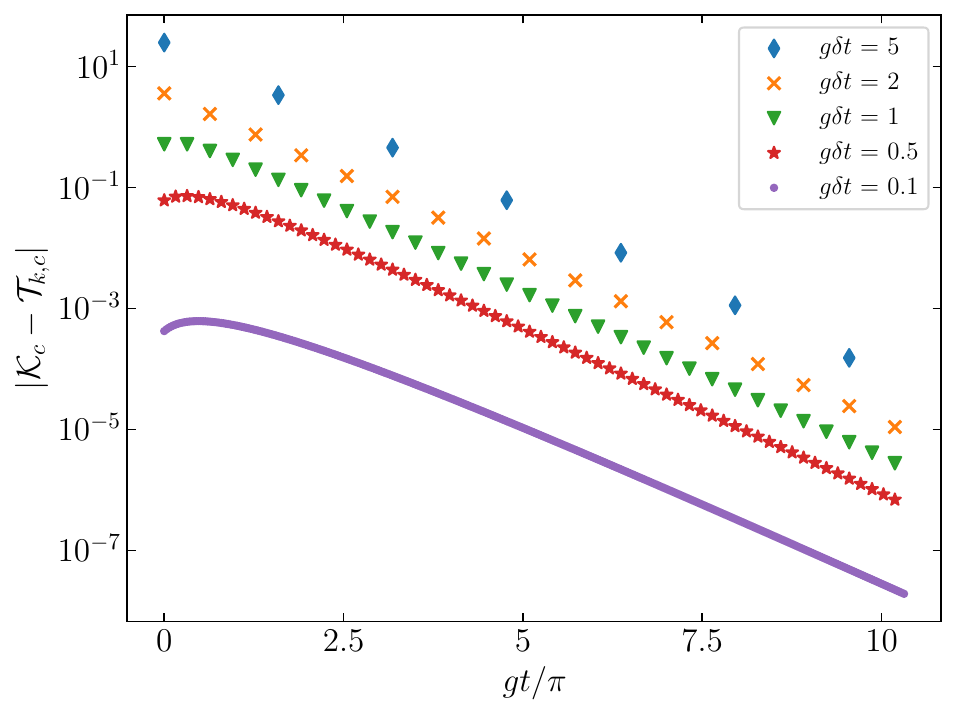}
    \caption{Absolute value of the difference between the memory kernel $\mK_c$ and the transfer tensor $\mT_{k,c}$ for different timestep values and $\kappa/4g = 0.2$, showing that $\mT_{k,c}$ converges to $\mK_c$ as the timestep is reduced.}
    \label{fig:K_vs_Tk}
\end{figure}

\begin{figure}
    \centering
    \includegraphics[width=0.95\linewidth]{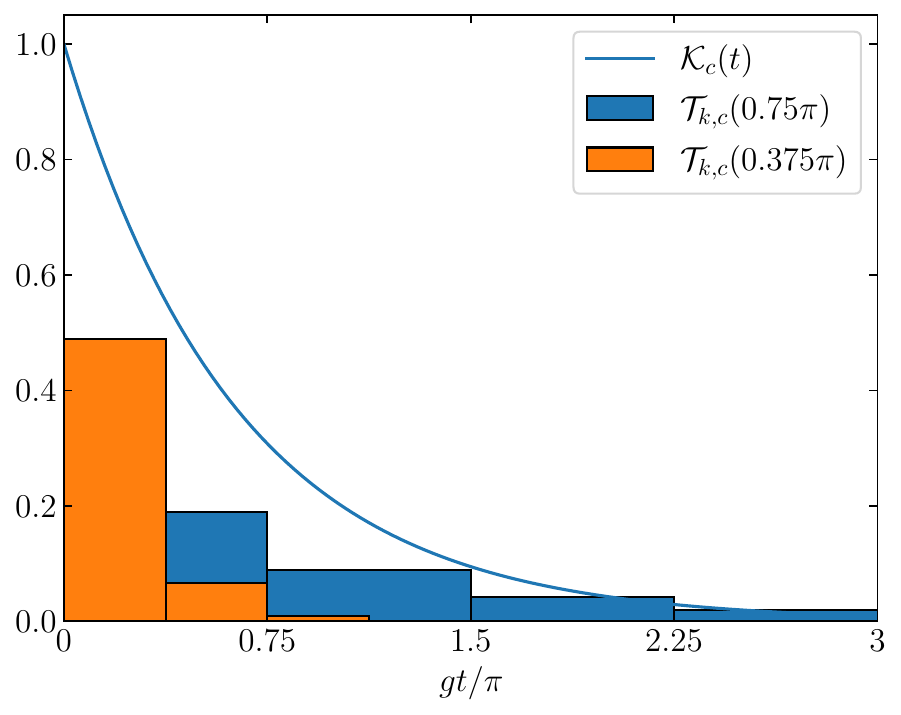}
    \caption{The solid line shows the decay of $\mK_c(t)$ as a function of time. The bars show the decay of $\mT_{k,c}$ as a function of time for two different time-step discretizations. Both the memory kernel and the TTs have been obtained for a value $\kappa/4g = 0.25$, corresponding to the underdamped regime of the system.}
    \label{fig:Kc_Tk}
\end{figure}

\begin{figure}
    \centering
    \includegraphics[width=\linewidth]{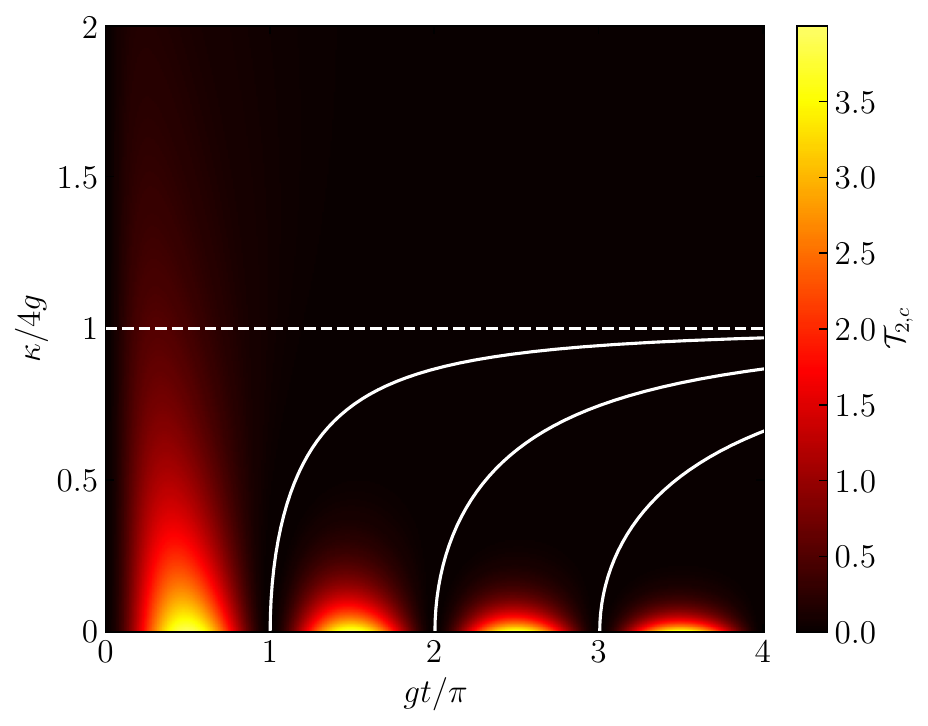}
    \caption{$\mathcal{T}_{2,c}$ as a function of $gt$ and $\kappa/4g$. The dashed line corresponds to the boundary between the underdamped and the overdamped regimes. The solid lines show the zeros of $\mT_{2,c}$ where the evolution of the system is guaranteed to be fully Markovian.}
    \label{fig:T2c_heatmap}
\end{figure}

\begin{figure}
    \centering
    \includegraphics[width=\linewidth]{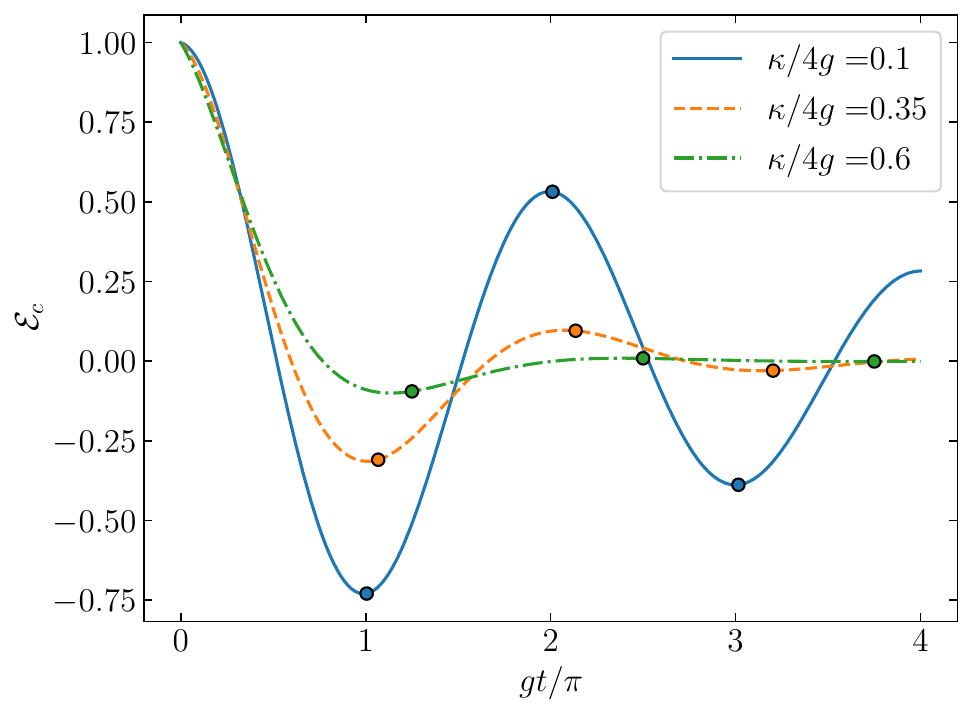}
    \caption{Temporal evolution of the coherence dynamical map $\mE_c$ in the underdamped regime. The dots on each line correspond to the three first zeros of $\mT_{2,c}(t)$, where the evolution is fully Markovian for that given timestep.}
    \label{fig:Ec_underdamped}
\end{figure}

To obtain the dynamical map and transfer-tensor components,
we perform the eigendecomposition of $\mL_B$. The eigenvalues are $\kappa_\pm
=
-{\kappa}/{4}
\pm
 \sqrt{({\kappa}/{4})^2 -  g^2 }$  and the projectors associated to their respective eigenvectors
\be
\mP_\pm=\frac{1}{\kappa_\mp-\kappa_\pm }\begin{pmatrix}
\kappa_\mp & ig \\
ig & -\kappa_\pm
\end{pmatrix}.
\ee
This provides $\mU_B=\mP_+e^{\kappa_+t}+\mP_-e^{\kappa_-t}$, which contains a component of the dynamical map 
\be
\mE_c(t)=\frac{\kappa_+e^{\kappa_- t}-\kappa_-e^{\kappa_+ t}}{\kappa_+ - \kappa_-}.
\ee
Depending on the quotient $\kappa/4g$, overdamped, critically damped and underdamped regimes may be distinguished in the dynamics of the coherence $c(t)=\mE_c(t)c(0)$, (see Fig.~\ref{fig:3_regimes}), associated with coherent-incoherent transitions, widely studied in the context of polaritonic dynamics \cite{WuCerrilloCao2024, qutubudinCao2025}.
We may now construct the second transfer tensor associated with the coherence,
\be
\mT_{2,c}\mP_B=\mP_B\mU_B\mQ_B\mU_B\mP_B=-\left(g\frac{e^{\kappa_- t}-e^{\kappa_+ t}}{\kappa_+ - \kappa_-}\right)^2\mP_B.
\ee
Together with the scalar $\mQ_B\mU_B\mQ_B$  this expression allows us to obtain all higher-order transfer-tensor components for the coherence sector:
\be
\label{eq:Tkc}
\mT_{k,c}(t)= \mT_{2,c}(t) \left(\frac{\kappa_+e^{\kappa_+ t}-\kappa_-e^{\kappa_- t}}{\kappa_+ - \kappa_-}\right)^{k-2}
\ee
One can verify that this result satisfies Eq.~\eqref{eq:lim}, as expected. This behavior is illustrated in Fig.~\ref{fig:K_vs_Tk}: as the time discretization becomes finer, the difference between the memory kernel $\mK_c(t)$ and the transfer tensor $\mathcal{T}_{k,c}$ correspondingly decreases.
Interestingly, while the memory kernel always decays monotonically (see Fig.~\ref{fig:Kc_Tk}), the transfer tensors retain explicit information about the dynamical regime of the system and decay over finite intervals due to their inherently discrete temporal structure.
Figure~\ref{fig:T2c_heatmap} illustrates this by showing $\mathcal{T}_{2,c}(t)$ for different values of the ratio $\kappa/4g$. 
In the underdamped regime, $\mathcal{T}_{2,c}(t)$ can be written as
\begin{equation}\label{eq:T2c}
\mathcal{T}_{2,c}(t)
    = 4g^{2} t^{2} e^{-\frac{\kappa t}{2}}
      \,\mathrm{sinc}^{2}\!\left[
          g t \sqrt{1 - \left(\frac{\kappa}{4g}\right)^{2}}
      \right],
\end{equation}
which vanishes whenever the argument of the sinc function is proportional to $\pi$. Since all coherence-sector transfer tensors scale with $\mathcal{T}_{2,c}$, as seen in Eq.~\eqref{eq:Tkc}, these zeros mark the timesteps at which concatenation of the dynamical map introduces no error. In other words, such points correspond to Markovian discretizations. Thus, full Markovianity arises precisely when the timestep matches the oscillation period of the coherences in the underdamped regime, as shown in Fig.~\ref{fig:Ec_underdamped}, where the zeros of $\mathcal{T}_{2,c}(t)$ align with the periodicity of the coherence dynamics.

\section{Population subspace}\label{sec:populations}
Subspace $A$ contains the three populations $p_e,p_i,p_g$ together with the imaginary part of coherence $c_{ei}$. Their evolution is governed by
\begin{equation}\label{eq:sistema_pops}
\left\lbrace\begin{array}{rcl}
\dot p_e&=&-2g c_{ei}^i,\\
\dot p_i&=&-\kappa p_i +2g c^i_{ei},\\
\dot p_g&=&\kappa p_i,\\
\dot c^i_{ei}&=&g (p_e-p_i)-\frac{\kappa}{2}c^i_{ei},
\end{array}\right.
\end{equation}
where $c_{ei}^i\equiv\text{Im}(c_{ei})$. 
It is convenient to introduce the relevant variable
$\delta p = p_\uparrow - p_\downarrow$ and the irrelevant variables
$c_{ei}^i$ and $\delta p' = p_i - p_g$. \textcolor{black}{In terms of these newly defined relevant and irrelevant variables, the system of equations \eqref{eq:sistema_pops} becomes}
$\dot{\mathbf{p}} = \mL_A \mathbf{p}$
with 
\[
\mathbf{p}=\begin{pmatrix}
\delta p \\
c^i_{ei}  \\
\delta p' \\
1
\end{pmatrix};\qquad
 \mL_A =\begin{pmatrix}
0 & -4 g & 0 & 0\\
3g/4 & -\kappa/2  & -g/2 & g/4 \\
\kappa/2 &  2 g & -\kappa  & -\kappa/2 \\
0 & 0 & 0 & 0
\end{pmatrix}.
\]
The projection superoperators in this subspace are $ \mP_A =1\oplus 0_{3\times3}$ and $ \mQ_A =0\oplus 1_{3\times3}$.

To obtain the memory kernel, it is necessary to exponentiate the matrix $\mQ_A\mL_A\mQ_A$. Following the same procedure as in Sec.~\ref{sec:coherences} we derive the corresponding expressions (see Appendix.~\ref{sec:AKp}) and find that it may be written in terms of dynamical objects of the coherence subspace:
\be
\mK_p(t)=\mK_c(t)\left[\mE_c(t)+2\frac{\mT_{3,c}}{\mT_{2,c}}\right].
\ee
Comparing the population component of the memory kernel with its coherence
counterpart shows that the population sector acquires a biexponential form,
with regimes of underdamping and critical damping.

For the dynamical map, 
the matrix $\mathcal{L}_A$ can be analytically
exponentiated, leading to a compact expression that connects the population
dynamics to quantities from the coherence subspace:
\be
\mE_p(t)=\mE_c^2(t) +\frac{1}{2}\mT_{2,c}.
\label{eq:Ep}
\ee
This shows that the population dynamics follows the coherence dynamics except
when non-Markovian effects - encoded in the second transfer tensor of the
coherence sector - become relevant. Using Eq.~\eqref{eq:Ep}, the second transfer
tensor of the populations is obtained as
\be
\mT_{2,p}=\frac 3 2 \mT_{2,c} \mE_p +\mT_{3,c} \mE_c + \frac{1}{2} \mT_{4,c},
\ee
as derived in Appendix.~\ref{sec:T2p_Appendix}. Importantly, $\mT_{2,p}$ remains proportional to $\mT_{2,c}$
and therefore inherits the same regimes of Markovian timesteps.

\begin{figure}[H]
    \centering
    \includegraphics[width=\linewidth]{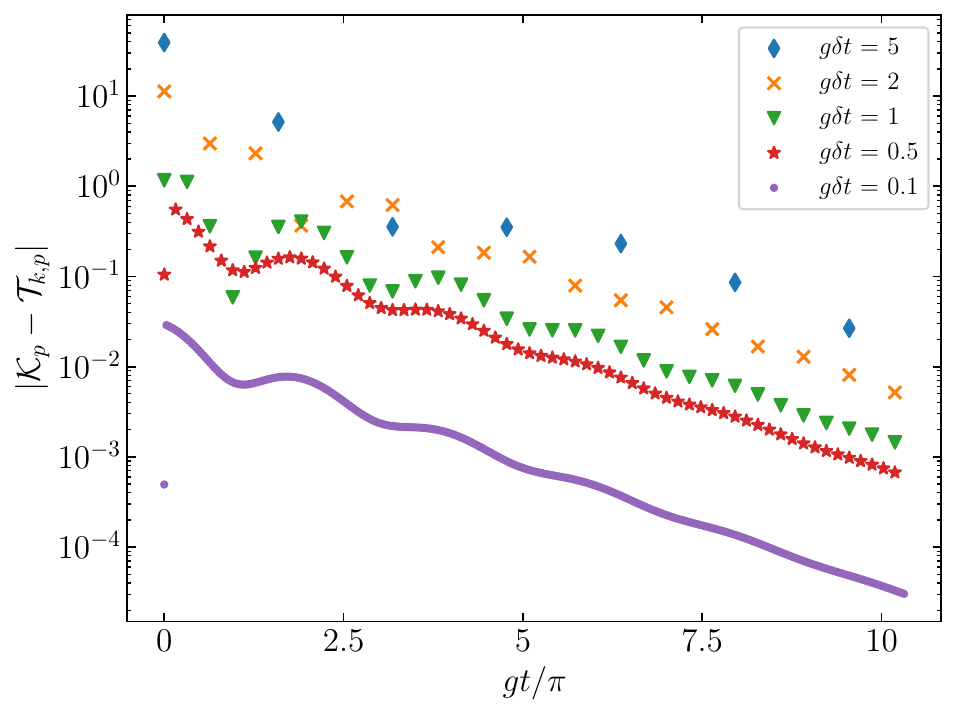}
    \caption{Absolute value of the difference between the memory kernel $\mK_p$ and the transfer tensor $\mT_{k,p}$ for different timestep values and $\kappa/4g = 0.2$, showing that $\mT_{k,p}$ converges to $\mK_p$ as the timestep is reduced.}
    \label{fig:K_vs_Tk_populations}
\end{figure}

Finally, the remaining transfer tensors can be obtained analytically by combining
Eq.~\eqref{eq:Ep} with the expressions derived for the coherence sector. As in the
case of the coherences, any finite discretization of the population memory kernel
$\mathcal{K}_p$ deviates from the exact transfer tensor $\mathcal{T}_{k,p}$.
Nevertheless, this deviation decreases as the timestep is reduced, in accordance
with Eq.~\eqref{eq:lim} and as illustrated in Fig.~\ref{fig:K_vs_Tk_populations}.

\section{Conclusion}\label{sec:conclusions}
We have demonstrated that the transfer tensors and the Nakajima-Zwanzig memory kernel are fundamentally distinct mathematical objects, converging only in the limit of an infinitesimal time-step. By applying this framework to the analytically solvable Jaynes-Cummings model, we have identified three distinct dynamical regimes associated with coherent-incoherent transitions via the cavity. Notably, our results indicate that Markovian behaviour is restricted to the underdamped regime.\textcolor{black}{ We further find that the dynamics of populations and coherences are linked through the structure of the second transfer tensor, which captures the leading-order deviation from Markovian composition at a finite time resolution. These findings highlight the role of the TTM as an exact, discrete-time framework to simulate non-Markovian systems beyond standard continuous-time characterisations.}

\textcolor{black}{Several questions remain open. It is natural to ask to what extent these features are generalizable beyond the Jaynes-Cummings model to more general open quantum systems and parameter regimes. In particular, the possible universality of transfer tensors being smaller in magnitude than the corresponding memory kernel, as well as the appearance fully Markovian behavior for certain time-steps, merit further investigation and will be addressed in future work.}

\begin{acknowledgments}
    M.M.R., A.L.G., G.R.R. and Javier C. acknowledge support from grant CNS2023-144994 funded by MICIU/AEI/10.13039/201100011033 and by ``ERDF/EU''. Javier C. additionally acknowledges support from European Union project C-QuENS (Grant No. 101135359). Jianshu Cao. acknowledges financial support from the MIT RSC grant and NSF No. CHE2324300.
\end{acknowledgments}

%\nocite{*}
\bibliography{ref} % Produces the bibliography via BibTeX.
\bigskip  % Adds vertical space

\newpage


%\bibliography{NVQ3} % Produces the bibliography via BibTeX.
\bigskip  % Adds vertical space

\appendix

\section{\textcolor{black}{General Expression for the TTs}}\label{sec:Tk_appendix}
\textcolor{black}{In order to provide a general expression for the $n$-th transfer tensor, we will resort to their definition}
\begin{equation}
    {\color{black}\mT_k = \mE_k - \sum_{j = 1}^{k}\mT_j\mE_{k-j}.}
\end{equation}
\textcolor{black}{Let's remind that the $k$-th dynamical map can be expressed as $\mE_k = \mP\mU^k\mP$, where we have made use of the projector $\mP$, which projects onto the relevant subspace. By definition $\mT_1 = \mE_1$. Therefore, to see the general form of the TTs, we have to start from $n=2$:}
\begin{equation}
     {\color{black}\mT_2 = \mE_2 - \mT_1\mE_1 = \mP\mU^2\mP - \mP\mU\mP^2\mU\mP,}
\end{equation}
\textcolor{black}{and by making use of the complementary projector $\mQ = \mathbb{I} - \mP$ and the idempotency property of projectors ($\mP^2 =\mP$), we can express $\mT_2$ as}

\begin{align}
     {\color{black}\mT_2 =} &  {\color{black}\mP\mU^2\mP - \mP\mU\mP^2\mU\mP}\notag \\
      {\color{black}=}&  {\color{black}\mP\mU(\mQ + \mP)\mU\mP - \mathcal{PUPUP} = \mathcal{PUQUP}}.
\end{align}
\textcolor{black}{Now for $\mT_3$:}
\begin{align}
    {\color{black}\mT_3 =} & {\color{black}\mE_3 - \mT_1\mE_2 - \mT_2\mE_1}\notag \\
     {\color{black}=}& {\color{black}\mathcal{PU}^3\mP - \mathcal{PUP}^2\mU^2\mP - \mathcal{PUQUP}^2\mathcal{UP}},
\end{align}
\textcolor{black}{which after some algebra becomes}
\begin{equation}
     {\color{black}\mT_3 = \mathcal{PUQ}(\mathcal{QUQ})\mathcal{QUP}.}
\end{equation}
\textcolor{black}{In the case of $\mT_4$:}
\begin{align}
     {\color{black}\mT_4}&  {\color{black} = \mE_4 - \mT_1\mE_3 - \mT_2\mE_2 - \mT_3\mE_1}\notag\\
      &  {\color{black}=\mathcal{PU}^4\mP - \mathcal{PUP}^2\mU^3\mP }\notag\\
      & {\color{black} - \mathcal{PUQUP}^2\mU^2\mP - \mathcal{PUQUQUP}^2\mathcal{UP},}
\end{align}
\textcolor{black}{which after algebraic manipulations becomes:}
\begin{equation}
    {\color{black} \mT_4 = \mathcal{PUQ}(\mathcal{QUQ})^2\mathcal{QUP}}.
\end{equation}
\textcolor{black}{One can iteravely construct higher order TTs and realize that the $k$-th TT can be expressed as}
\begin{equation}
    {\color{black} \mT_k = \mathcal{PUQ}(\mathcal{QUQ})^{k-2}\mathcal{QUP}.}
\end{equation}
\section{Reducing TTM to Memory Kernel}\label{sec:TTM_K_lim}

Given a homogeneous time discretization where the timestep is $\delta t = t/k$, where $k$ is the number of time-steps, the calculation of the limit of the TTM to the memory kernel only for small timesteps ($\delta t\to 0$, or equivalently $k\to \infty$) is as follows
\begin{align}
    \lim_{\delta t\to 0}\frac{\mT_k\left(\delta t\right)}{(\delta t)^2} =& \lim_{k\rightarrow\infty}\frac{\mT_k\left(t/k\right)}{t^2/k^2} \notag\\ =&\lim_{k\rightarrow\infty}\mP\left(1+\frac{t}{k}\mL\right)\left[\mQ\left(1+\frac{t}{k}\mL\right)\right]^{k-1}\mP\notag \\
    =&\mP\mL\mQ\lim_{k\rightarrow\infty}\left(1+\frac{t}{k}\mQ\mL\mQ\right)^{k-2}\mQ\mL\mP
\end{align}

The remaining limit is the definition of the exponential
\be
\lim_{k\rightarrow\infty}\left(1+A\frac{t}{k}\mQ\mL\mQ\right)^{k-2}=e^{At}.
\ee
up to a correction of order $k^{-2}$.

\section{Eigensolution of Subspace B}

Simplified expression for the coherence, where the associated memory kernel appears
\be
\dot c(t)=-g^2\int_0^te^{-\kappa(t-s)/2}c(s)ds.
\ee
For the transfer tensors, the full exponential of $\mL_B$ is needed.
In order to compute it, we will use its spectral decomposition. The eigenvalues of
\be
 \mL_B =\begin{pmatrix}
0 & -i g \\
-i g & -{\kappa}/{2},
\end{pmatrix}
\ee
 are the solutions to the characteristic polynomial
\be
\lambda^2+\kappa\lambda/2+g^2=0,
\ee
which read
\be
\kappa_\pm=-\frac{\kappa}{4}\pm\frac{\sqrt{\kappa^2/4-4g^2}}{2}.
\ee
The corresponding right eigenvectors read
\be
\mathbf v_\pm=\begin{pmatrix}
-ig \\
\kappa_\pm
\end{pmatrix},
\ee
and the left eigenvectors are simply $\mathbf w_\pm=\mathbf v_\pm^T$ due to the fact that $\mL_B = \mL_B^T$. The associated projectors are obtained as
\be
\mP_\pm=\frac{\mathbf v_\pm\mathbf w_\pm}{\mathbf w_\pm \mathbf v_\pm}=\frac{1}{\kappa_\pm^2-g^2 }\begin{pmatrix}
-g^2 & -ig\kappa_\pm \\
-ig\kappa_\pm & \kappa_\pm^2
\end{pmatrix}
\ee
which simplify to
\be
\mP_\pm=\frac{1}{\kappa_\mp-\kappa_\pm }\begin{pmatrix}
\kappa_\mp & ig \\
ig & -\kappa_\pm
\end{pmatrix}.
\ee
as a consequence of the invariance of the determinant $\kappa_+\kappa_-=g^2$.

\section{\label{sec:AKp} Populations Memory Kernel}

To compute the populations memory kernel, we need to compute the exponential matrix $\exp(\mQ_A\mL_A\mQ_At)$, which will be done by using the eigendecomposition of that matrix. The eigenvalues of
\be
\mQ_A\mL_A\mQ_A=0\oplus \begin{pmatrix}
 -\kappa/2  & -g/2 & g/4 \\
 2 g & -\kappa  & -\kappa/2 \\
0 & 0 & 0 
\end{pmatrix}.
\ee
The first block has eigenvalue 0, and the eigenvalues of the $3\times3$ block are the solution to the polynomial
\be
-\lambda\left[\left(\frac \kappa 2+\lambda\right)\left(\kappa+\lambda\right)+g^2\right]=0,
\ee
which read $\lambda_0=0$ and
\be
\lambda_\pm=-\frac{3\kappa}{4}\pm\frac{\sqrt{\kappa^2/4-4g^2}}{2}=-\frac \kappa 2 +\kappa_\pm
\ee
and produce the right eigenvectors
\be
\mathbf v_\pm=\begin{pmatrix}
\kappa_\mp\\
-2g\\
0
\end{pmatrix},\qquad 
\mathbf v_0=\begin{pmatrix}
 \kappa g  \\
g^2-\kappa^2/2\\
2g^2+\kappa^2
\end{pmatrix},
\ee
and left eigenvectors
\be
\mathbf w^T_\pm=\begin{pmatrix}
\kappa_\mp\\
g/2\\
\frac g 2 \frac{\kappa_\mp-\kappa}{2\kappa_\pm-\kappa}
\end{pmatrix},\qquad 
\mathbf w^T_0=\begin{pmatrix}
0  \\
0\\
1
\end{pmatrix},
\ee
\\
which are the solution to $\mL_A\mathbf{w}^T_\lambda = \lambda\mathbf{w}^T_\lambda$.
The associated projectors for the $3\times 3$ block are
\be
\mP_0=\frac{\mathbf v_0\mathbf w_0}{\mathbf w_0 \mathbf v_0}=\frac{1}{\kappa^2+2g^2 }\begin{pmatrix}
0&0&\kappa g  \\
0&0&g^2-\kappa^2/2\\
0&0&2g^2+\kappa^2
\end{pmatrix},
\ee
and
\be
\mP_\pm=\frac{\mathbf v_\pm\mathbf w_\pm}{\mathbf w_\pm \mathbf v_\pm}=\frac{1}{\kappa_\mp-\kappa_\pm }\begin{pmatrix}
\kappa_\mp&g/2&  \frac {g} 2 \frac{\kappa_\mp-\kappa}{2\kappa_\pm-\kappa}\\
-2g&-\kappa_\pm&-\kappa_\pm\frac{\kappa_\mp-\kappa}{2\kappa_\pm-\kappa}\\
0&0&0
\end{pmatrix},
\ee
from which the memory kernel can be extracted as
\be
\mK_p(t)=\mK_c(t)\frac{(\kappa_+-2\kappa_-)e^{\kappa_- t}-(\kappa_--2\kappa_+)e^{\kappa_+ t}}{\kappa_+ - \kappa_-}.
\ee

\section{Eigensolution of Subspace A}
We consider the matrix $\mL_A$
\be
 \mL_A =\begin{pmatrix}
0 & -4 g & 0 & 0\\
3g/4 & -\kappa/2  & -g/2 & g/4 \\
\kappa/2 &  2 g & -\kappa  & -\kappa/2 \\
0 & 0 & 0 & 0
\end{pmatrix}.
\ee
Its characteristic polynomial
\be
\frac{1}{2}\gamma(2\lambda + \kappa)(4g^2 + \gamma[\gamma + \kappa]) = 0,
\ee
yields eigenvalues $\gamma_0=0$, $\gamma_1=-\kappa/2$ and $\gamma_\pm=2\kappa_\pm$ and the associated projectors are
\be
\mP_0=\begin{pmatrix}
0 & 0 & 0 & -1\\
0 & 0 & 0 & 0 \\
0 & 0 & 0 & -1 \\
0 & 0 & 0 & 1
\end{pmatrix},
\ee
\be
\mP_1=\frac{-g}{(\kappa_+-\kappa_-)^2}\begin{pmatrix}
g & -2\kappa & 2g & 3g\\
\kappa/8 &  -\kappa^2/4g & \kappa/4 & 3\kappa/8 \\
3g/2 & -3\kappa & 3g & 9g/2 \\
0 & 0 & 0 & 0
\end{pmatrix},
\ee
\\
\vspace{0.5cm}
\begin{widetext}
\be
\mP_\pm=\frac{-\kappa_\pm}{(\kappa_+-\kappa_-)^2}\begin{pmatrix}
\kappa_\mp/2-\kappa_\pm & -4 g & -\kappa_\mp & (\kappa_\mp+\kappa)/2\\
 g(1/2 -\kappa_\mp/4\kappa_\pm)& 2\kappa_\mp  & -g\kappa_\mp/2\kappa\pm &   g(1/2 +\kappa_\mp/4\kappa_\pm)\\
\kappa/4(1-\kappa_\mp/\kappa_\pm)-3\kappa_\mp/4 & -2g(1+2\kappa_\mp/\kappa_\pm)  & \kappa\kappa_\mp/2\kappa_\pm -2\kappa_\mp & (2g^2+\kappa^2)/8\kappa_\pm \\
0 & 0 & 0 & 0
\end{pmatrix}.
\ee
\end{widetext}

\section{Derivation of $\mathcal{T}_{2,p}$}\label{sec:T2p_Appendix}
Starting from the general expression for the $k$-th transfer tensor $\mT_k$ from Eq.~\eqref{eq:Tk}, it is immediate to see that the $\mT_2$ is obtained as
\begin{equation}
    \mT_{2,p} = \mP_A\mU_A\mQ_A\mU_A\mP_A.
\end{equation}
Using the fact that $\mP_A,\mQ_A$ are projectors, we can express $\mQ_A = 1-\mP_A$ and rewrite the $\mT_{2,p}$ as
\begin{equation}\label{eq:T2p_subspaceA}
    \mT_{2,p} = \mP_A\mU_A\mU_A\mP_A - \mP_A\mU_A\mP_A\mU_A\mP_A.
\end{equation}
Recalling that the dynamical map of the populations $\mE_p = \mP_A\mU_A\mP_A$ can be expressed as in Eq.~\eqref{eq:Ep}, we can state that 

\begin{align}
    \mE_p = & \mP_A\mU_A\mP_A\notag\\
    =&\mE^2_c + \frac{1}{2}\mT_{2,c} = \mP_B\mU_B\mP_B\mU_B\mP_B + \frac{1}{2}\mP_B\mU_B\mQ_B\mU_B\mP_B\notag\\
    =&\mP_B\mU_B\left(\mP_B + \frac{1}{2}\mQ_B\right)\mU_B\mP_B.
\end{align}
Using this expression we can rewrite Eq.~\eqref{eq:T2p_subspaceA} as
\begin{widetext}
    \begin{align}
        \mT_{2,p} = & \mP_A\mU_A\mU_A\mP_A - \mP_A\mU_A\mP_A\mU_A\mP_A = \notag\\
        =&\mP_B\mU_B\left(\mP_B + \mQ_B\right)\mU_B\left(\mP_B + \frac{1}{2}\mQ_B\right)\mU_B\left(\mP_B+\mQ_B\right)\mU_B\mP_B - \mP_B\mU_B\left(\mP_B + \frac{1}{2}\mQ_B\right)\mU_B\mP_B\mU_B\left(\mP_B + \frac{1}{2}\mQ_B\right)\mU_B\mP_B\notag\\
        =&\frac{3}{2}(\mP_B\mU_B\mP_B\mU_B\mP_B)(\mP_B\mU_B\mQ_B\mU_B\mP_B) + \frac{3}{4}(\mP_B\mU_B\mQ_B\mU_B\mP_B)^2 \notag\\ 
        &+ (\mP_B\mU_B\mQ_B\mU_B\mQ_B\mU_B\mP_B)(\mP_B\mU_B\mP_B) + \frac{1}{2}(\mP_B\mU_B\mQ_B\mU_B\mQ_B\mU_B\mQ_B\mU_B\mP_B)\notag\\
        =&\frac{3}{2}\mE^2_c\mT_{2,c} + \frac{3}{4}\mT^2_{2,c} + \mT_{3,c}\mE_c + \frac{1}{2}\mT_{4,c} = \frac{3}{2}\mT_{2,c}\mE_p +  \mT_{3,c}\mE_c + \frac{1}{2}\mT_{4,c}.
    \end{align}
\end{widetext}
These derivations have used the fact that the components of both the transfer tensors and the dynamical maps are scalars.
\end{document}